\documentclass{emulateapj}

\begin{document}


\title{On Asymmetric Distributions of Satellite Galaxies}


\author{A. Bowden\altaffilmark{1}, N.W. Evans\altaffilmark{1}, V. Belokurov\altaffilmark{1}}
\affil{Institute of Astronomy, University of
  Cambridge, Madingley Road, Cambridge CB3 0HA, UK}
\email{nwe@ast.cam.ac.uk}



\begin{abstract}
We demonstrate that the asymmetric distribution of M31 satellites
cannot be produced by tides from the Milky Way as such effects are too
weak. However, loosely bound associations and groups of satellites can
fall into larger haloes and give rise to asymmetries. We compute the
survival times for such associations. We prove that the survival time
is always shortest in Keplerian potentials, and can be $\sim3$ times
longer in logarithmic potentials. We provide an analytical formula for
the dispersal time in terms of the size and velocity dispersion of the
infalling structure.  We show that, if an association of $\sim 10$
dwarfs fell into the M31 halo, its present aspect would be that of an
asymmetric disk of satellites. We also discuss the case of cold
substructure in the Andromeda II and Ursa Minor dwarfs.
\end{abstract}




\section{Introduction}

Asymmetric distributions of satellite galaxies are seemingly common.
\citet{Mc06} found that all bar one of the known 16 satellites of
M31 lie on the hemisphere of the M31 sky facing the Milky Way
Galaxy.  The more recent investigation of \citet{Co13} using a larger
sample has soothed worries that this was an artefact of small sample
size or observational bias. \citet{Ib13} analyzed the radial
velocities of the satellites and argued that 13 of the satellites
possess a coherent rotational motion, with 12 of the 13 lying on the
side of M31 nearest to the Milky Way Galaxy.

For our Galaxy, there are also tantalizing hints of a similar
asymmetric distributions, though our `fishbowl' viewpoint makes such
claims hard to evaluate. Plots of the locations of the known Milky Way
satellites in the sky seem to suggest that the northern Galactic
hemisphere is over-endowed compared to the southern. Most of the
northern Galactic hemisphere lies on the far side of the Milky Way as
judged from an M31 perspective. In other words, the Milky Way
satellites are also asymmetrically distributed with a preponderance on
the far side from M31.

What causes such asymmetries in satellite distributions to be set up
and how do they maintain themselves? At first sight, tidal forces
suggest themselves as a possibility. Suppose the Milky Way and M31
have mass $M$ and separation $D$ in the $x$-direction, then the tidal
potential in the frame of the host is
\begin{equation}
\Phi_{\rm tidal} = -\frac{GM}{\sqrt{(D-x)^2 + y^2 + z^2}}.
\end{equation}
Using a Taylor expansion in the limit $x,y,z \ll D$ for a distant
perturber, the tidal force is
\begin{equation}
F_x = \frac{GM}{D} \left(\frac{1}{D} + \frac{2x}{D^2} + \frac{3x^2}{D^3} - \frac{3y^2}{2D^3} - \frac{3z^2}{2D^3}\right).
\end{equation}
The first term here is $GM/D^2$. As this calculation is done in the
frame of the host galaxy, this cancels with the acceleration due to
the non-inertial frame. We then have several terms which are linear in
$x$ and these cannot lead to an asymmetry in the satellite
population. The lowest order term that can generate any asymmetry is
the $3GMx^2/D^4$ term. The magnitude of this effect for a Milky
Way-M31 pair is roughly $\sim 1$ kpc, as can be verified by
simulations in the center of mass frame of the Milky Way and M31. This
is clearly insufficient to explain the asymmetry in the population of
M31 satellites.

This leads us to consider the possibility that such asymmetries arise
from the initial conditions of infall of satellites. Significant
fractions of satellites may be accreted from a similar direction in
groups, or in loosely bound associations or clumps and this can lead
to asymmetries in the satellite
distributions~\citep[e.g.,][]{Li08,Do08}. Motivated by the example of
M31, \citet{Sh13}, \citet{Go13} and \citet{Sa14} provide scenarios in
which groups of satellite galaxies may be accreted in preferred
directions.  \citet{Li11} demonstrate that in simulations of the Local
Group, satellite infall is not spherical in nature and that preferred
directions are observed. \citet{Ba14} show that planar features are
not uncommon in cosmological simulations, however the features they
describe are transient in nature. The extent to which such
associations and phase space structures can persist over a Hubble time
without dispersal through phase-mixing is not immediately an obvious.

\begin{figure*}
 \begin{center}
  \includegraphics[scale=0.27]{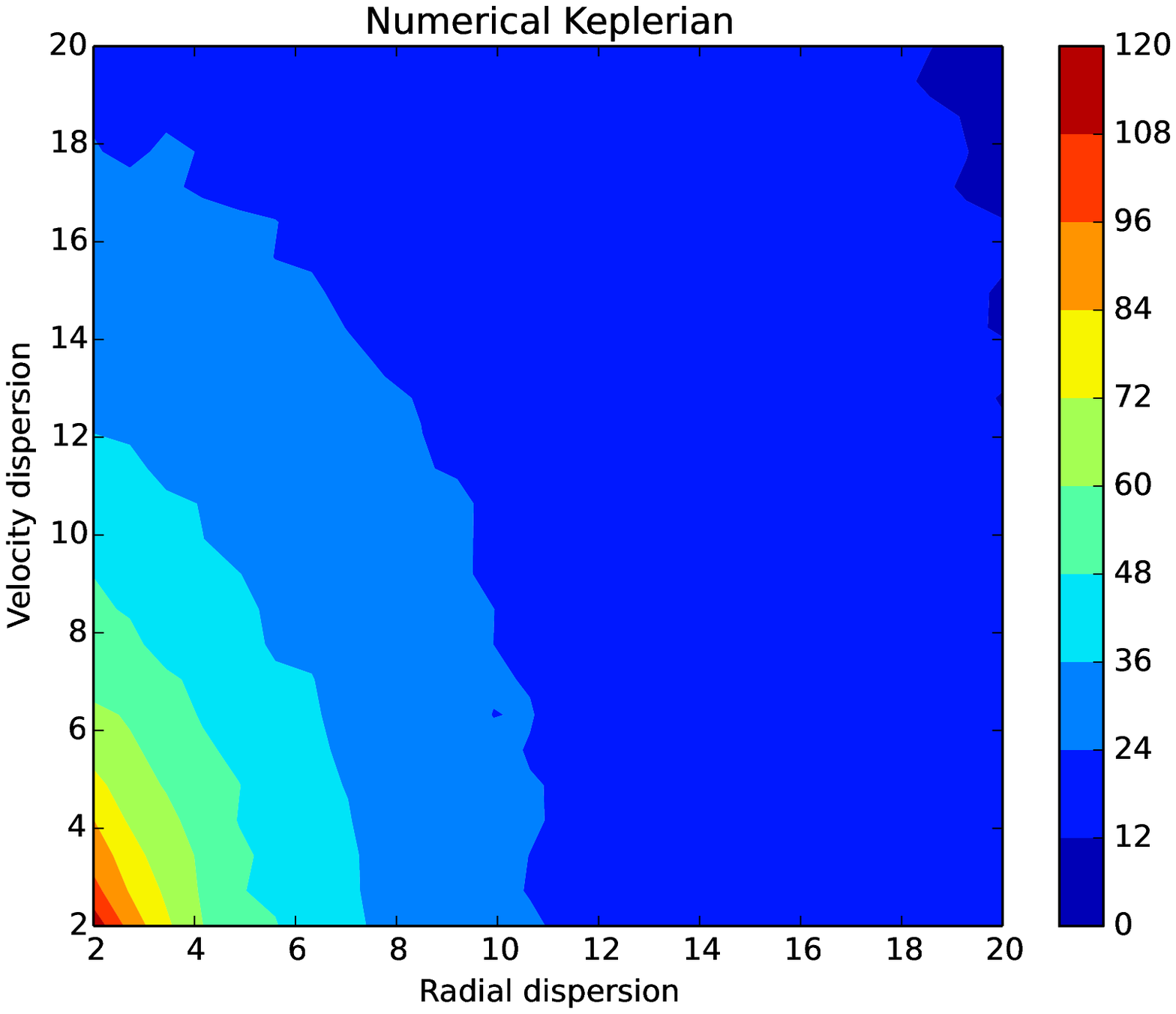}
  \includegraphics[scale=0.27]{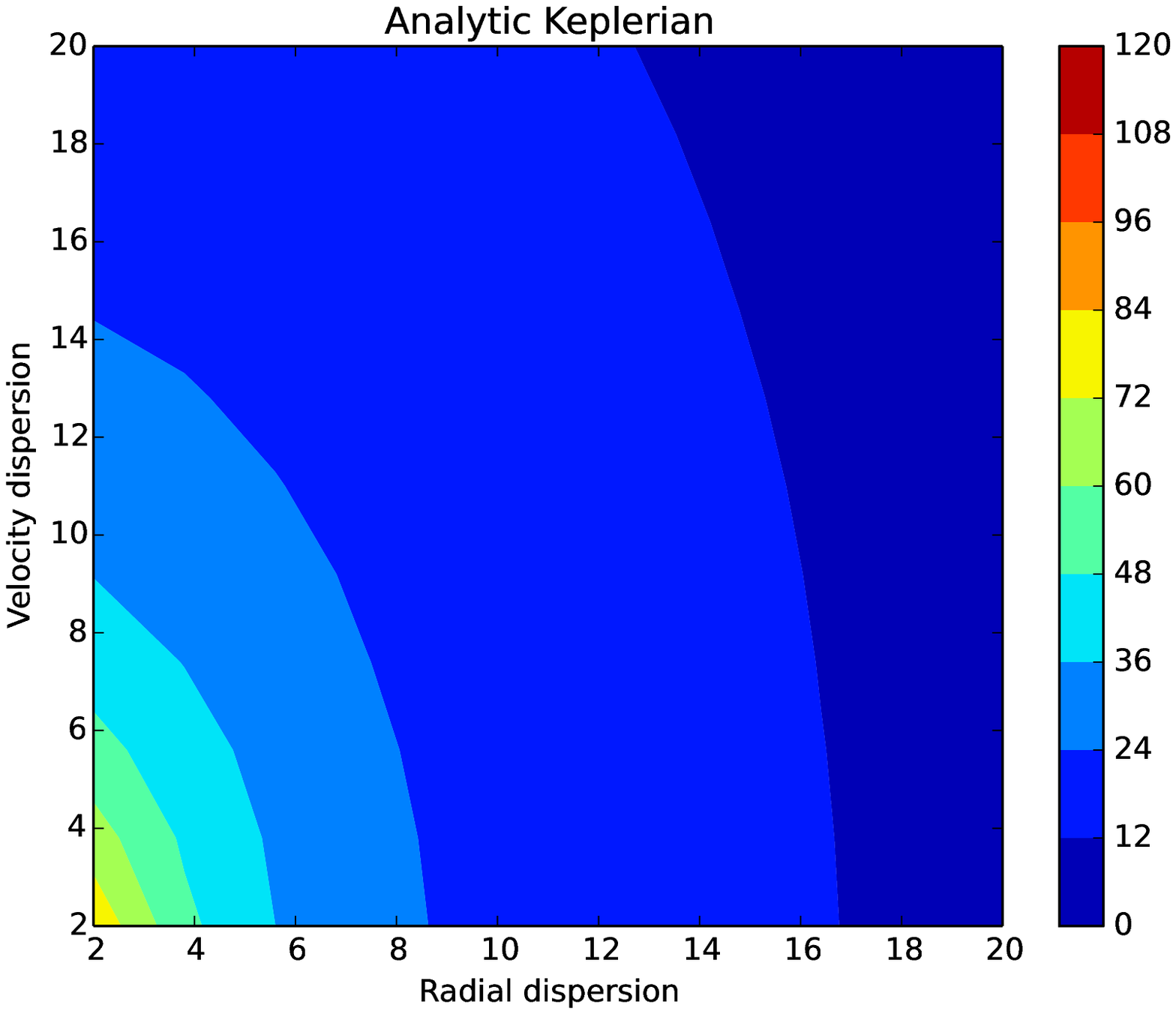}
  \includegraphics[scale=0.27]{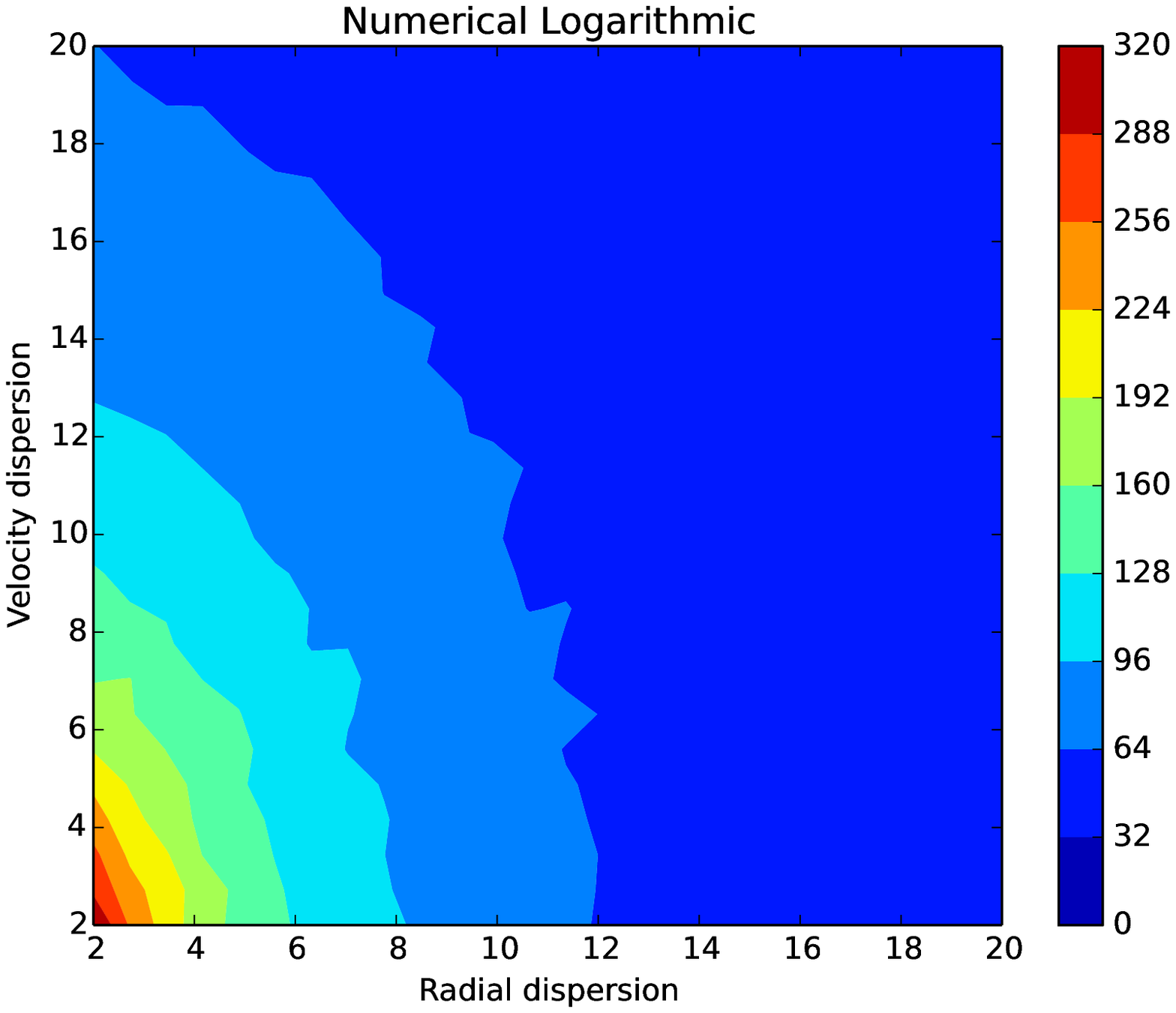}
 \end{center}
\caption{Contour plots showing the time taken for the rms $\Delta\phi$
  to reach $2\pi$ as a function of dispersion in position $\sigma_r$
  and velocity $\sigma_v$ of the Gaussian clump of
  eq.~(\ref{eq:gaussclump}) for (left, middle panels) Keplerian and
  (right panel) logarithmic potentials. The left and right panels have
  been computed by direct simulation, whilst the middle panel uses the
  analytic formula of eq.~(\ref{eq:analytic}).  All the panels refer
  to the case of a 10$^{10}M_\odot$ association, which is placed on an
  orbit with $\mu_r = 200$ kpc and $\mu_v = 150$ kms$^{-1}$.  This is
  tailored for the case of dispersal of an assemblage of dwarf
  galaxies in the outer parts of the halo of a galaxy like M31.}
\label{fig:numvsana}
\end{figure*}

\section{The Dispersal Theorem}

An observable asymmetry may indicate that an initially clumpy
population has not had sufficient time to fully phase mix. This
suggests that we examine the dispersal of such clumps.  The most
natural coordinate system in which to do so is action-angles. Here,
actions are conserved while angles evolve linearly with
time~\citep[see e.g.,][]{Go80}
\begin{equation}
 \theta = \theta_0 + \omega t.
\end{equation}
Let us consider the dispersal of a clump in action-angle space. For
our application, the clump is composed of a loose association of
satellite galaxies, though the theorem we prove holds more
generally. As the clump disperses, the spread in actions (which are
adiabatic invariants) is conserved, whilst the spread in angles is
\begin{equation}
 \Delta\theta = \delta\theta_0 + \delta\omega t.
\end{equation}
So, the dispersal of the clump in time therefore only depends on
$\delta\omega$, the difference in frequencies as actions are changed
by a small amount, that is, the Hessian matrix multiplied by the
difference in actions. This is similar to the
related problem in stream dynamics~\citep[c.f.,][]{Sa13}.

We can compute the Hessian for scale-free power-law
potentials~\citep{Ev94}, which take the form:
\begin{equation}
\Phi(r) = Ar^{\alpha},\qquad A = {v_0^2\over \alpha r_0^\alpha}.
\label{eq:powlaw}
\end{equation}
Here, $v_0$ is the circular velocity at radius $r_0$.  When $\alpha
=0$, this becomes the isothermal sphere, whilst when $\alpha = -1$
this is of course the Kelperian case. The outer parts of galaxy haloes
lie between these two limiting cases.

In a spherical potential, there are two actions, namely the radial
action $J_r$ and the azimuthal action $J_\phi$ which may be taken as
the angular momentum $L$~\citep[see e.g.,][]{Go80}.  The Hamiltonian
as a function of the actions takes the form~\citep{Wi14}
\begin{equation}
 H(L,J_r) = C (J_\phi + D J_r)^\beta,
\end{equation}
Here, $\beta = 2\alpha/(\alpha+2)$ and the constants $C$ and $D$
are given in Williams et al (2014).
%
%
Differentiation gives the angular frequency as
\begin{equation}
 \omega_\phi = (A\alpha)^{2/(\alpha+2)}(J_\phi + D J_r)^{(\alpha-2)/(\alpha+2)}.
\end{equation}
One more differentiation gives the components of the Hessian
\begin{equation}
 \frac{\partial^2H}{\partial J_\phi\partial J_i} = (A\alpha)^{2/(\alpha+2)}
 \frac{\alpha-2}{\alpha+2} a_i (J_\phi + D J_r)^{-4/(\alpha+2)},
\end{equation}
where $a_i$ is $D$ for $J_r$ and $1$ for $J_\phi$. This is a
decreasing function of $\alpha$ if the power-law models are normalised
to the same enclosed mass. {\it So, for a fixed enclosed mass, the closer
the potential is to Keplerian, the larger the magnitude of the Hessian
and the faster the dispersal.} This is the dispersal theorem.

We may be interested in the number of orbits $n_{\rm orb}$ it takes
for a clump to spread out. This means we divide out the frequency
$\omega_\phi$ by the Hessian matrix,
\begin{equation}
n_{\rm orb} \propto \omega_\phi (\frac{\partial^2H}{\partial
  J_\phi\partial J_i})^{-1} = \frac{\alpha+2}{\alpha-2}
\frac{(J_\phi+DJ_r)}{a_i}.
\label{eq:norb}
\end{equation}
As $D$ increases as a function of $\alpha$, it is straightforward to
establish that the complete expression is an increasing function of
$\alpha$. In other words, the Keplerian case is again the most
efficient at mixing.

An assumption in these calculations is that the spread in angles
dominates the evolution of the clump, whilst the spread in actions is
small. This assumption is likely to hold good for small clumps.
Therefore, it is important to test the results against simulations,
which we proceed to do in the next section.

\section{The Dispersal of Groups of Satellites}

\subsection{Keplerian Case}

To describe the degree of phase mixing as a clump disperses, we use
the mean angular separation between galaxies in the association. This
can be computed both analytically and numerically. We consider phase
mixing to be complete when this value reaches $2\pi$.

To start with, we assume the simplest case of a Keplerian galactic
host potential. In this potential, orbits have periods given by
\begin{equation}
P_{\rm orb} = \frac{\pi GM}{\sqrt{2}} \epsilon^{-3/2},
\end{equation}
where $\epsilon$ is the energy and $M$ is the mass of the host galaxy.
Provided the orbit is not highly eccentric, then the angular phase is
\begin{equation}
\phi \approx \frac{2\pi t}{P_{\rm orb}} = \frac{2\sqrt{2}}{GM} \epsilon^{3/2} t.
\end{equation}
The magnitude of the difference in $\phi$ between two such satellites
is
\begin{equation}
(\Delta\phi)^2 = A (\epsilon_1^{3/2} - \epsilon_2^{3/2})^2 = A (\epsilon_1^{3} + \epsilon_2^{3} - 2\epsilon_1^{3/2}\epsilon_2^{3/2}),
\end{equation}
where $A = 8t^2/(G^2M^2)$.  To find the mean, we integrate over phase
space co-ordinates after multiplying by a probability distribution. We
model the group of satellites as a (truncated) Gaussian with
dispersions in position and velocity $\sigma_r$ and $\sigma_v$ around
means $\mu_r$ and $\mu_v$
\begin{eqnarray}
P &=& \frac{1}{(2\pi\sigma_r\sigma_v)^2} \exp-\bigg(
\frac{(r_1 - \mu_r)^2}{2\sigma_r^2} \\ &+& \frac{(r_2 -
  \mu_r)^2}{2\sigma_r^2} + \frac{(v_1 - \mu_v)^2}{2\sigma_v^2} +
\frac{(v_2 - \mu_v)^2}{2\sigma_v^2} \bigg).\nonumber
\label{eq:gaussclump}
\end{eqnarray}
We thus evaluate numerically
\begin{equation}
\overline{(\Delta\phi)^2} = A \int P (\epsilon_1^{3} + \epsilon_2^{3}
- 2\epsilon_1^{3/2}\epsilon_2^{3/2}) dr_1 dr_2 dv_1 dv_2.
\label{eq:analytic}
\end{equation}
truncating our probability distribution at $4\sigma$.

\begin{figure}
 \begin{center}
\includegraphics[width=.45\textwidth]{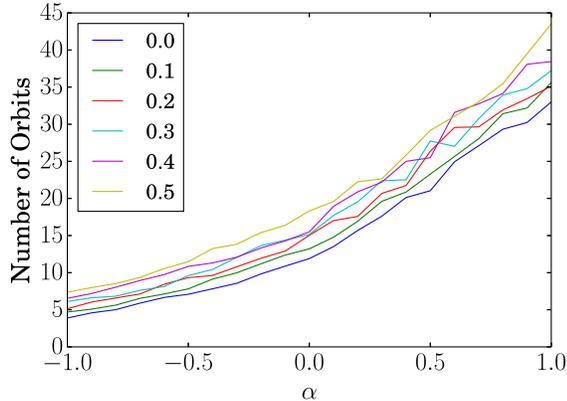}
 \end{center}
\caption{Plot showing the number of orbits required for a Gaussian
  clump to phase mix as a function of power-law index for scale-free
  spherical potentials. The mean position and rotational velocity of
  the clump is the same as in Fig.~1. Different lines correspond to
  different orbital eccentricities for the clump. We observe that the
  efficiency of phase mixing is an increasing function of power-law
  index $\alpha$, with the Keplerian case as the fastest
  dispersing. This vindicates the dispersal theorem.}
\label{fig:fixapoorbs}
\end{figure}
\vskip1.5cm

\subsection{Logarithmic Case}

For arbitrary spherical potentials, such as the flat rotation curve or
logarithmic case, we must evaluate
\begin{equation}
\overline{(\Delta\phi)^2} = C \int P \left(\frac{1}{T^2_{\phi_1}} +
\frac{1}{T^2_{\phi_2}} - \frac{1}{T_{\phi_1}T_{\phi_2}}\right) dr_1
dr_2 dv_1 dv_2.
\end{equation}
Here, $T_\phi$ is the azimuthal period, given by $T_\phi = 2\pi
T_r/|\Delta\phi|$, with
\begin{equation}
 T_r = 2 \int^{r_2}_{r_1} \frac{dr}{\sqrt{2[E - \Phi(r)] - L^2/r^2}},
\end{equation}
and
\begin{equation}
 \Delta\phi = 2L \int^{r_2}_{r_1} \frac{dr}{r^2\sqrt{2[E - \Phi(r)] - L^2/r^2}}.
\end{equation}
The validity of these approximations can be tested via direct
comparison to numerical simulations.  The simulations involve drawing
test particles from a 6-D Gaussian centered on the mean position and
velocity. These particles are integrated forwards in time in the host
galactic potential, assuming no self-gravity, and their mean angular
separation evaluated.

Fig.~\ref{fig:numvsana} shows the time taken in the Keplerian and
logarithmic cases for the mean $\Delta\phi$ to reach $2\pi$, for a
range of values of $\sigma_r$ and $\sigma_v$. In each case, the clump
is centered on an orbit with $\mu_r = 200$ kpc and $\mu_v = 150$
kms$^{-1}$, as might be appropriate for the outer parts of a large
galaxy like M31. The enclosed mass of the host galaxy is $M = 2.2
\times 10^{12} M_\odot$, comparable to estimates of the mass of
M31~\citep[see e.g.,][]{Ev00, Di14}. The analytic approximation
reproduces well the shape of the contours, however it slightly
underestimates the time taken for a clump to disperse.  The likely
sources of the discrepancy are the approximations we have made,
particularly the assumption that the angular velocity of the particles
is constant in time, allowing us to perform our integrals in terms of
the angular period. Further, we modeled our probability distribution
as Gaussian in the magnitude $r$ and $v$. This requires the implicit
assumption that our perturbations about the mean clump velocity are in
the direction of motion, and perturbations about the position are
radial. In actuality, we should perform an integral over the 6-D phase
space. However, perturbations to the relevant quantities or $r$ and
$v$ from a six-dimensional Gaussian are dominated by the contributions
from the radial position terms and the direction of motion velocity
terms.

Fig.~\ref{fig:fixapoorbs} shows, for the same clump mean velocities,
the number of orbits required for the mean $\Delta \phi$ to reach
$2\pi$ for a variety of power-law potentials as given in
eq~(\ref{eq:powlaw}).  This demonstrates that the number of orbital
periods needed for dispersal is an increasing function of $\alpha$, as
suggested by our calculation in Section 2. Note that the simulations
have really pushed beyond the frozen action approximation used
there. Comparing the Keplerian and Logarithmic potentials -- which are
the likely bounding cases for the outer parts of galaxy haloes -- an
identical clump takes more orbital periods to disperse in the
logarithmic halo by a factor of $\sim 3$.

One assumption underlying all the work so far is that the disk of
satellites lies in a nearly spherical potential. This is a natural
choice, as orbits in spherical potentials lie in a plane. However, the
disk of satellites could lie in the equatorial plane of an
axisymmetric potential, or in one of the two stable principal planes
of a triaxial potential~\citep{Bo13}. We have verified by numerical
calculations that the dispersal time and the number of orbits for
$\Delta \phi$ to reach $2\pi$ is close to the spherical case in these
instances.

Whilst our numerical simulations are simple in nature, they display
qualitatively similar results to more complex simulations such as those
in \citet{Li11}, where Local Group substructure is shown to persist over
long timescales.


\section{Applications to Galactic Substructure}
In order to apply our results to the expected survival timescales of
observed substructure, we desire a simple analytic formula for
estimating a lower limit (Keplerian case) for the time taken for a clump
of satellites to disperse. From eq~(\ref{eq:norb}), we see that
\begin{equation}
 n_{\rm orb} \propto \frac{J}{|\delta J|},
\end{equation}
where $\Delta J$ is the spread in the sum of the actions. Using our
expression for the Hamiltonian, we can show
\begin{equation}
 \delta J = \frac{GM \delta\epsilon}{2\sqrt{2}\epsilon^{\frac{3}{2}}},
\end{equation}
where $\epsilon$ is reduced energy. Using
\begin{equation}
  \delta\epsilon = -\frac{GM\delta r}{r^2} - v\delta v,
\end{equation}
we fit the functional form for the dispersal time
\begin{equation}
  T \propto n_{\rm orb}P_{\rm orb} = A\frac{(GM)^{\frac{2}{3}} P_{orb}^{\frac{1}{3}}}
 {\sqrt{ (\frac{GM \sigma_r}{ \mu^2_r})^2 + (\mu_v \sigma_v)^2}}.
\end{equation}
To determine the parameter $A$, we simultaneously fit five different
orbits each with a 10x10 grid of $\sigma_r$ between $2-20$ kpc and
$\sigma_v$ between $2-20$ kms$^{-1}$. The values were calculated
analytically using the method described in Section 3.1.  The orbital
parameters for $\mu_r$ /kpc, $\mu_v$ /kms$^{-1}$ were
$[200,150]$,$[100,250]$,$[100,180]$,$[150,180]$,$[150,150]$.  Using
Monte Carlo methods, we recover a best fit value of $A = 0.738$.  The
mean absolute fractional error for the fit is $0.065$, and the maximum
is $0.326$. The high maximum errors occurred in cases two and three,
when $\mu_r = 100$ kpc and $\sigma_r = 20$ kpc. This is when we expect
our approximations to break down -- the Taylor expansion for
calculating $\delta\omega$ is no longer valid in this regime. We only
expect our formula to be applicable in the case when $\Delta J/J$ is
much less than unity, ruling out for example the dispersion of
structure on highly radial orbits.

\subsection{The M31 satellites}

One application of this formula is for the dispersal of a clump of
satellites falling into M31. We take an M31 mass of $2 \times 10^{12}
M_{\odot}$. Let us assume that an association of dwarfs of total mass
of $10^{10} M_\odot$ fell into the M31 halo and was the progenitor of
$\sim 10$ dwarf galaxies in the present-day disc of satellites.  Note
that the total mass in the association is rather modest, as it is
smaller by a factor of $\sim 10$ as compared to nearby associations of
dwarf galaxies studied observationally by \citep{Tu06}.  We take
$\sigma_r$ as $20$ kpc, which, for a loosely bound clump with
potential energy shape factor of $~0.3$, fixes $\sigma_v$ at $20$
kms$^{-1}$.  In order to mimic the properties of the M31 galaxies, we
place the association on an orbit with apocenter at $200$ kpc and
pericentre at $100$ kpc. This gives an apocenter velocity of $~170$
kms$^{-1}$. In these units, this orbit has a period of $3.93$
kpckm$^{-1}$s. Our analytic formula then gives a lower limit of $~10$
Gyr for the clump to disperse. This is of order the age of the universe,
and we would expect this timescale to be roughly 3 times larger in a
logarithmic halo.

In other words, if an association of $\sim 10$ dwarfs galaxies was
accreted by M31 sometime within the last $~10$ Gyr, then we expect the
association to persist rather easily and the present-day distribution
would retain memory of the initial conditions. The satellites would
not be fully phase-mixed and would be spread out to lie in an
asymmetric disk, much as observed.

\subsection{Substructure in And II and UMi}

Another application is to the cold substructure in dwarf spheroidal
galaxies, such as Andromeda II and Ursa Minor. These are probably the
the result of the engulfment of smaller dwarfs and clusters within the
larger dwarf spheroidal itself.

The cold stellar stream in \citet{Am14} is at a distance of roughly
$1.3$ kpc with a dispersion of order $0.1$ kpc. The velocity
dispersion is given as less than $3$ kms$^{-1}$. The enclosed mass
interior to the stream is $2.5$x$10^8 M_{\odot}$. This gives a
circular velocity at $1.3$ kpc of $28.8$ kms$^{-1}$. This orbit has a
period of $0.3$ kpc km$^{-1}$s. A lower limit to the dispersal
timescale for these values is $450$ Myr. Even for a logarithmic halo,
this would suggest that, if there is a detectable angular asymmetry in
the substructure, then it is relatively young. The data at this time
certainly indicate that the angular extent of the stream is not a full
$2\pi$ around Andromeda II, although this may be partly a consequence
of the spectroscopic selection function.

The cold clump in Ursa Minor was discovered by~\citet{Kl03}, who
suggested that it may be a cluster that had fallen into the dwarf
galaxy and was in the process of dissolving. As the velocity offset
between the cold clump and Ursa Minor is small, this suggests that the
orbit is either radial or is circular and viewed almost face-on.
\citet{Kl03} investigated the former possibility and showed that the
dissolution timescale depended on the properties of the central parts
of the potential. Here, we follow the latter possibility, namely that
that cold substructure is on a near circular orbit at 150 pc from the
center. The mass enclosed within the orbit is $1.25 \times 10^7
M_\odot$. The size and the velocity dispersion of the clump are given
by \citet{Kl03} as $0.012$ pc and $0.5$ kms$^{-1}$ respectively, which
gives a phase-mixing timescale of $\sim 130$ Myr. This is comparable
to the disruption timescale in a cusped potential found by
\citet{Kl03} if the clump is on a radial orbit. Note however we have
probably pushed our dispersal formula to beyond its formal domain of
applicability in studying disruption of such a tightly bound clump.

\section{Conclusions}

Our main conclusion is a general formula for the timescale for
phase-mixing of orbits in nearly spherical potentials.  Survival times
are always shortest in Keplerian potentials and can be $\sim$ 3 times
longer in logarithmic potentials. We use the formula to demonstrate
that the asymmetric distribution of the disk of satellites of M31 can
be maintained without undue difficulty in nearly spherical
potentials. This is because the outer parts of such galaxies have
enormously long memories.  If a loose association of dwarf galaxies
were accreted together, then the substructure can survive for
timescales longer than the age of the Universe. The accreted dwarfs
would still show appreciable spatial asymmetries, much as the M31
satellites do.

\acknowledgments 
AB thanks STFC for financial support. We thank the referee for a
careful and thoughtful reading of the manuscript.

\end{document}